\documentclass{article}
\usepackage{ijcai25}

\usepackage{times}
\usepackage{soul}
\usepackage{url}
\usepackage[hidelinks]{hyperref}
\usepackage[utf8]{inputenc}
\usepackage[small]{caption}
\usepackage{graphicx}
\usepackage{amsmath}
\usepackage{amsthm}
\usepackage{amssymb}
\usepackage{booktabs}
\usepackage{algorithm}
\usepackage{algorithmic}
\usepackage[switch]{lineno}
\usepackage{multirow}
\usepackage{tabularx}
\usepackage{enumitem}

\title{Causal Learning for Trustworthy Recommender Systems: A Survey}

\author{
Jin Li$^1$
\and
Shoujin Wang$^1$\and
Qi Zhang$^2$\and
Longbing Cao$^3$\and
Fang Chen$^1$\and\\
Xiuzhen Zhang$^4$\and
Dietmar Jannach$^5$\And
Charu C. Aggarwal$^6$
\affiliations
$^1$University of Technology Sydney, \ 
$^2$Tongji University, \\
$^3$Macquarie University, \ 
$^4$RMIT University, \\
$^5$University of Klagenfurt, \
$^6$IBM T. J. Watson Research Center
}

\begin{document}

\maketitle

\begin{abstract}
Recommender Systems (RS) have significantly advanced online content filtering and personalized decision-making. However, emerging vulnerabilities in RS have catalyzed a paradigm shift towards Trustworthy RS (TRS). Despite substantial progress on TRS, most efforts focus on data correlations while overlooking the fundamental causal nature of recommendations. This drawback hinders TRS from identifying the root cause of trustworthiness issues, leading to limited fairness, robustness, and explainability. To bridge this gap, causal learning emerges as a class of promising methods to augment TRS. These methods, grounded in reliable causality, excel in mitigating various biases and noise while offering insightful explanations for TRS. However, there is a lack of timely and dedicated surveys in this vibrant area. This paper creates an overview of TRS from the perspective of causal learning. We begin by presenting the advantages and common procedures of Causality-oriented TRS (CTRS). Then, we identify potential trustworthiness challenges at each stage and link them to viable causal solutions, followed by a classification of CTRS methods. Finally, we discuss several future directions for advancing this field.
\end{abstract}

\section{Introduction}
Recommender Systems (RS), as one of the most successful applications of artificial intelligence, have revolutionized the way of our decision-making in the digital world. However, with their growing influence, concerns over trustworthiness of RS have become increasingly prevalent. Inspecting the \textbf{lifecycle} of RS, i.e., data preparation, representation learning, recommendation generation, and evaluation,  \cite{TRS_TIST}, we identify several significant trustworthiness challenges, which are the obstacles that hinder the system's fairness, robustness and explainability. They present pivotal yet challenging objectives for modern RS, extending beyond accuracy. Specifically, fairness requires unbiased decision-making for various users, items, and providers. Robustness emphasizes the reliability of RS against noise or attacks. Explainability poses an urgent need for strong interpretability, advancing the understanding of RS mechanisms and results. Recently, causal learning has emerged as a powerful tool for addressing Trustworthy RS (TRS) challenges \cite{DBLP:journals/tois/GaoWLCHLLZJ24,DBLP:conf/icml/LiXZ0023}, leveraging causality rather than mere correlations. In this paper, we overview the trustworthiness challenges throughout the RS lifecycle, alongside a comprehensive investigation of causality-oriented solutions.

\begin{figure}[t]
  \centerline{\includegraphics[width=0.42\textwidth]{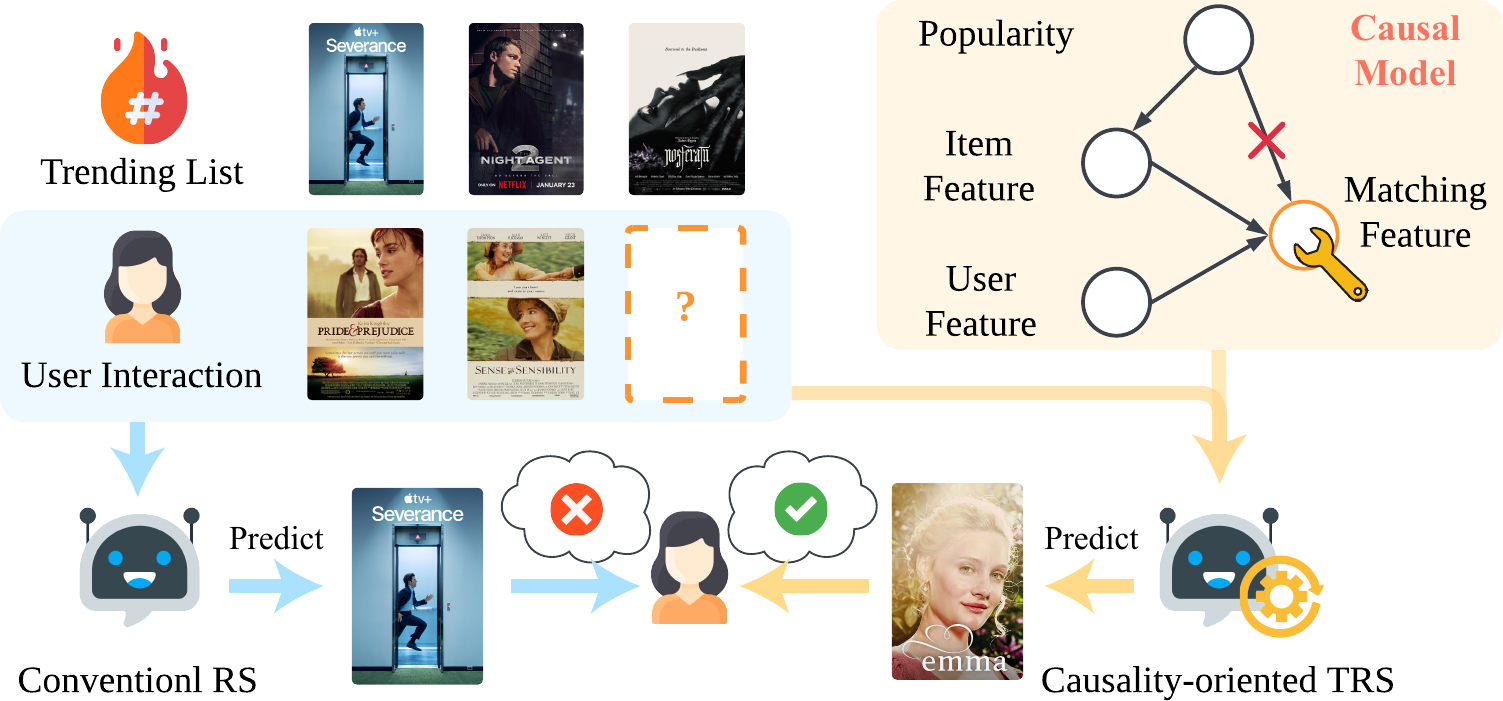}}
  \caption{Conventional RS relies on correlations, often misled by popularity bias (e.g., suggesting trending thrillers to users who prefer period dramas). CTRS performs causal interventions to reduce bias, ensuring more relevant and trustworthy recommendations.}
  \label{fig::CTRS}\vspace{-10pt}
\end{figure} 

\subsubsection{Motivation: why Causality-oriented TRS (CTRS)?}
\noindent\textbf{Causality is more reliable than correlations.}
Conventional RS heavily relies on correlations learned from observed data. However, correlations are not always causally correct, and spurious correlations can be notably misleading, as exemplified in Figure \ref{fig::CTRS}. In contrast, CTRS methods \cite{DBLP:journals/tois/00010FSYL023} are adept at identifying genuine causal relations and applying causal adjustments for enhanced trustworthiness. Moreover, another persistent issue in RS is the non-random nature of missing data \cite{DBLP:conf/icml/LiXZ0023}, as user-item interactions are typically preference-driven. Causal techniques like propensity reweighting \cite{DBLP:conf/www/LiXZ023} can help mitigate this issue, leading to more accurate and reliable recommendations.

\noindent\textbf{Causal learning offers theoretical instructions for identifying and eliminating biases.}
Causal learning provides various theoretically grounded approaches to analyze the bias in RS via straightforward yet insightful causal models. Upon targeting the true cause of biases, a wide range of causal techniques (e.g., structure adjustment and counterfactual inference) can be applied to measure and adjust the underlying cause-and-effect relationships. Benefiting from this, a significant enhancement in the fairness of RS can be achieved.

\noindent\textbf{Causal learning helps provide a straightforward and effective explanation.}
RS models, especially deep learning-based ones, often lack interpretability, and thus can hardly offer clues about the reason for recommendations. In contrast, CTRS leverages path discovery and counterfactual inference for superior interpretability. These methods \cite{DBLP:conf/sigir/XianFMMZ19,DBLP:journals/tois/WangLYLX24} not only illuminate recommendation mechanisms but also clarify the rationale behind recommendation results, thereby providing more transparent and comprehensible frameworks for both users and developers.

\subsubsection{How does causal learning help solve TRS problems?}
Causal learning enhances TRS by explicitly modeling, analyzing, and intervening in the underlying cause-and-effect relationships. Specifically, it operates through two key steps: causal formulation and causal inference. The first step delineates causal relationships within the system by constructing causal models \cite{Rubin1974Estimating}, such as Structural Causal Models (SCM) \cite{Pearl2009Causality}. It identifies relevant variables (e.g., user preferences, item attributes) and maps their causal influences, either through causal discovery or domain expertise. In the second step, causal techniques are applied to estimate and adjust for causal effects. Thus, we can move beyond correlation-based models to understand and manipulate the causal factors affecting trustworthiness for robust, fair, and explainable recommendations. 

\subsubsection{Comparison with Existing Surveys}
Despite several surveys on relevant topics, three critical gaps remain: 1) \textbf{Limited scope}: TRS surveys \cite{DBLP:journals/corr/abs-2209-10117,DBLP:journals/corr/abs-2207-12515,TRS_TIST} primarily review general aspects of trustworthiness in RS, such as fairness, robustness, and explainability. However, they lack a particular focus on causality-oriented TRS, which is a significant branch of TRS and cannot be overlooked. 2) \textbf{Coarse-grained challenge analysis}: Existing literature \cite{DBLP:journals/corr/abs-2209-10117,DBLP:journals/tist/LiCXGTLZ23} typically examines TRS challenges at a high level from general aspects, without mapping them to specific stages of RS lifecycle. 
Hence, it lacks insights into building a practical and unified TRS platform, as trustworthiness concerns vary across different stages in real-world applications. 3) \textbf{Lack of causal alignment}: While surveys on causal inference based RS \cite{DBLP:journals/corr/abs-2208-12397,DBLP:journals/corr/abs-2303-11666,DBLP:journals/corr/abs-2301-00910,DBLP:journals/corr/abs-2301-04016,DBLP:conf/ijcai/WuLDHDDSZZ22} focus on technical methodologies, they fail to explicitly align causal learning with trustworthiness challenges or provide insights on why and how causal learning can help build TRS.

\noindent\textbf{Contributions.} To bridge these gaps, we provide a dedicated survey on CTRS to examine trustworthiness challenges and corresponding causal solutions throughout the RS lifecycle. The main contributions are summarized as follows:
\begin{itemize}[leftmargin=*]

\item \textbf{Granular, stage-wise challenge analysis}: We introduce an in-depth examination and a novel taxonomy of key challenges in building TRS. Through well mapping different aspects of challenges to different stages when building RS, we offer a structured overview and actionable insights for addressing specific trustworthiness issues at each RS stage.

\item \textbf{Comprehensive CTRS method summary and comparison}: 
We provide an in-depth survey on CTRS methods, highlighting the unique advantages of causal learning in tackling TRS challenges. In addition, a critical comparison of CTRS methods is provided to guide the selection of appropriate solutions for addressing specific challenges.

\item \textbf{Insightful challenge-method alignment}: We highlight the underlying connections and alignment between trustworthiness challenges and causal learning methods, explaining why and how different causal learning methods address corresponding challenges. In addition, we offer valuable perspectives on future directions for advancing CTRS.
\end{itemize}

\begin{table*}[ht]
\centering
\scriptsize
\renewcommand{\arraystretch}{0.95}
\begin{tabularx}{\textwidth}{>{\hsize=0.9\hsize}X|>{\centering\arraybackslash\hsize=0.7\hsize}X|>{\centering\arraybackslash\hsize=0.9\hsize}X|>{\centering\arraybackslash\hsize=1.4\hsize}X|>{\centering\arraybackslash\hsize=1.5\hsize}X|>{\hsize=0.6\hsize}X}
  \toprule[1pt]
  \multicolumn{1}{c|}{RS stage} &Aspect &Sub-aspect &Trustworthiness challenge &Causal solution &\multicolumn{1}{c}{Example} \\ \midrule
  \multirow{8}{=}[-18pt]{Data preparation} &\multirow{6}{*}[-13pt]{Fairness} &\multirow{2}{*}[-3pt]{User side} &User attribute bias &Counterfactual inference &\scriptsize [1], [2] \\ \cmidrule{4-6}
  & & &Population bias &Counterfactual augmentation &- \\ \cmidrule{3-6}
  & &\multirow{4}{*}[-8pt]{Item side} &Item attribute bias &Counterfactual inference &\scriptsize [1], [2] \\ \cmidrule{4-6}
  & & &\multirow{2}{*}[-3pt]{Popularity bias} &Structure adjustment &\scriptsize [3], [4] \\ \cmidrule{5-6}
  & & & &Counterfactual inference &\scriptsize [5] \\ \cmidrule{4-6}
  & & &Position bias &Propensity reweighting &\scriptsize [6] \\ \cmidrule{2-6}
  &\multirow{2}{*}[-3pt]{Robustness} &Non-randomness &Selection bias &Propensity reweighting &\scriptsize [7], [8], [9] \\ \cmidrule{3-6}
  & &\multirow{1}{*}{Corruption} &Poisoning attacks &Counterfactual augmentation &- \\ \midrule
  \multirow{4}{=}[-8pt]{Representation learning} &Fairness &Implicit intention &Conformity bias &Causal disentanglement &\scriptsize [10] \\ \cmidrule{2-6}
  &\multirow{3}{*}[-5pt]{Robustness} &\multirow{3}{*}[-5pt]{Noise} &\multirow{2}{*}[-3pt]{Spurious correlations} &Structure adjustment &\scriptsize [11] \\ \cmidrule{5-6}
  & & & &Counterfactual augmentation &\scriptsize [12], [13] \\ \cmidrule{4-6}
  & & &Noisy feedback &Counterfactual inference &\scriptsize [14] \\ \midrule
  \multirow{4}{=}[-8pt]{Recommendation generation} &\multirow{2}{*}[-3pt]{Fairness} &\multirow{2}{*}[-3pt]{Model side} &Bias amplification &Structure adjustment &\scriptsize [15] \\ \cmidrule{4-6}
  & & &Filter bubble bias &Counterfactual inference &\scriptsize [16], [17] \\ \cmidrule{2-6}
  &\multirow{2}{*}[-3pt]{Explainability} &Model agnostic &Unexplainable results &Counterfactual inference &\scriptsize [18], [19], [20] \\ \cmidrule{3-6}
  & &Model specific &Black-box mechanism &Path discovery &\scriptsize [21] \\ \midrule
  \multirow{1}{*}{Evaluation} &Robustness &Model effect &Uplift evaluation &Uplift effect estimation &\scriptsize [22] \\ \midrule
  \multicolumn{6}{>{\hsize=\dimexpr 6\hsize+12\tabcolsep+\arrayrulewidth\relax}X}{\scriptsize \textsuperscript{[1]}\cite{DBLP:conf/sigir/LiCXGZ21}; \textsuperscript{[2]}\cite{DBLP:journals/tois/ChenWZZHXX24}; 
  \textsuperscript{[3]}\cite{DBLP:conf/sigir/ZhangF0WSL021}; \textsuperscript{[4]}\cite{DBLP:conf/cikm/GuptaSMVS21}; \textsuperscript{[5]}\cite{DBLP:conf/kdd/WeiFCWYH21}; \textsuperscript{[6]}\cite{DBLP:conf/ijcai/JoachimsSS18}; \textsuperscript{[7]}\cite{DBLP:conf/nips/Liu0023}; \textsuperscript{[8]}\cite{DBLP:conf/icml/LiXZ0023}; \textsuperscript{[9]}\cite{DBLP:conf/www/LiXZ023}; \textsuperscript{[10]}\cite{DBLP:conf/www/ZhengGLHLJ21}; \textsuperscript{[11]}\cite{DBLP:journals/tois/00010FSYL023}; \textsuperscript{[12]}\cite{DBLP:conf/sigir/MuLZWDW22}; \textsuperscript{[13]}\cite{DBLP:conf/ictir/XuGLFCZ23}; \textsuperscript{[14]}\cite{DBLP:conf/sigir/ZhangJSWXW21}; \textsuperscript{[15]}\cite{DBLP:conf/kdd/WangF0WC21}; \textsuperscript{[16]}\cite{DBLP:conf/sigir/WangFNC22}; \textsuperscript{[17]}\cite{DBLP:journals/tois/GaoWLCHLLZJ24}; \textsuperscript{[18]}\cite{DBLP:conf/cikm/TanXG00Z21}; \textsuperscript{[19]}\cite{DBLP:journals/tois/WangLYLX24}; \textsuperscript{[20]}\cite{DBLP:journals/tkde/WangLYLX24}; \textsuperscript{[21]}\cite{DBLP:conf/sigir/XianFMMZ19}; \textsuperscript{[22]}\cite{DBLP:conf/recsys/SatoSTSZO19}
  } \\
  \bottomrule[1pt]
\end{tabularx}
\caption{Trustworthiness challenges in the lifecycle of RS and corresponding causal learning solutions.}
\label{tab::classification}\vspace{-10pt}
\end{table*}

\section{Trustworthiness Challenges in RS Lifecycle}
\subsection{Problem Formulation}
Recommender systems aim to predict user preferences over items to provide personalized recommendations. Formally, let $\mathcal{U}=\{u_1,u_2,\cdots,u_M\}$ denote a set of $M$ users, and $\mathcal{I}=\{i_1,i_2,\cdots,i_N\}$ denote a set of $N$ items. The observed interactions between users and items are represented as $\mathcal{D} = \{(u,i,r_{ui})\mid u\in \mathcal{U}, i\in \mathcal{I}\}$, where $r_{ui}$ is the explicit rating or implicit feedback from user $u$ on item $i$. The goal of an RS is to learn a predictive function $f$, parameterized by $\Theta$, with user and item representations $\mathbf{h}_{u}$ and $\mathbf{h}_i$, such that $\hat{r}_{ui} = f(\mathbf{h}_u,\mathbf{h}_i\mid \Theta)$ estimates the true preference.

To achieve this goal, a typical RS comprises four successive and inter-connected stages, i.e., \textbf{data preparation}, \textbf{representation learning}, \textbf{recommendation generation}, and \textbf{evaluation}. Each stage plays a fundamental role in the RS lifecycle \cite{TRS_TIST} and is confronted with its own trustworthiness challenges. We highlight that the overall trustworthiness of RS essentially hinges on the reliability of each stage. Only when each stage is deemed trustworthy can we develop a unified trustworthy RS. In this section, we delve into the trustworthiness challenges inherent at each stage, examining them through the lens of causal learning. Table \ref{tab::classification} summarizes these challenges across the various stages and illustrates their alignment with causal solutions, offering a structured overview of the landscape.

\subsection{Challenges in Data Preparation}
As the first stage of an RS, data preparation involves gathering, organizing and transforming input data $\mathcal{D}$ (e.g., user-item interactions and side information). However, raw data collected from complex cyberspace often contain various biases (e.g., popularity bias), noise (e.g., fake ratings/reviews from online water armies), and even malicious attacks, posing significant risks to the system's fairness and robustness.

\subsubsection{Fairness Challenges in Data Preparation}
During data preparation, a primary fairness challenge lies in how to mitigate the pre-existing biases \cite{DBLP:journals/tist/LiCXGTLZ23} in raw data $\mathcal{D}$. Such inherent biases directly affect RS models trained on these data, potentially harming the interests of stakeholders. According to different subjects of fairness concerns, there exist two major types of pre-existing biases, i.e., user-side \cite{DBLP:conf/sigir/LiCXGZ21} and item-side bias \cite{Li2025Generating}. 

User-side bias refers to unfair recommendations for different groups of users. The first prevalent bias in this category is known as \textbf{user attribute bias}. Let $A_u$ denote a sensitive attribute associated with user $u$, such as gender, age, or ethnicity. User attribute bias occurs when RS models make discriminatory decisions based on these sensitive attributes, i.e.,
\begin{equation}
  \mathbb{E}[\hat{r}_{ui}\mid A_u=a] \neq \mathbb{E}[\hat{r}_{ui}\mid A_u=b],
\end{equation}
for different values $a$ and $b$ of $A_u$. For example, a biased job recommender may make discriminatory decisions based on the gender of users \cite{DBLP:conf/flairs/MansouryASDPM20}. Another critical user-side bias, namely \textbf{population bias}, stems from the unbalanced distribution of user data. Define subsets of users based on their activity levels, we have heavy users $\mathcal{U}_{\rm heavy}$ who are frequently active and casual users $\mathcal{U}_{\rm casual}$ who interact sporadically. Observe that $\mathcal{U}_{\rm heavy}$ typically contributes a significantly larger volume of interaction data than $\mathcal{U}_{\rm casual}$, i.e., $|\mathcal{D}_{\rm heavy}| \gg |\mathcal{D}_{\rm casual}|$, where $\mathcal{D}_{\rm heavy} = \{(u,i,r_{ui})\mid u\in\mathcal{U}_{\rm heavy}\}$ and $\mathcal{D}_{\rm casual} = \mathcal{D} \setminus \mathcal{D}_{\rm heavy}$. Thus, RS models trained on such data tend to favor heavy users and perform noticeably better for them than for casual users \cite{DBLP:conf/www/WangLCWCC22}.

Apart from user-side bias, in a trustworthy recommendation ecosystem \cite{TRS_TIST}, mitigating item-side bias helps to safeguard the interests of items and their providers. Similar to users, items may suffer from \textbf{attribute bias} due to certain attributes $A_i$ (e.g., brands and production places). However, beyond general attributes, the popularity of an item can markedly affect the recommendation results, giving rise to \textbf{popularity bias} \cite{DBLP:journals/umuai/KlimashevskaiaJET24}. Popular items $\mathcal{I}_{\rm pop}$, benefiting from their fame, naturally draw more attention than long-tail ones $\mathcal{I}_{\rm long-tail}$. The bias is shown as:
\begin{equation}
  \mathbb{E}[\hat{r}_{ui}\mid i\in\mathcal{I}_{\rm pop}] > \mathbb{E}[\hat{r}_{ui}\mid i\in\mathcal{I}_{\rm long-tail}],
\end{equation}
even when users do not necessarily prefer popular items. Another classic item-side bias, namely \textbf{position bias} \cite{DBLP:conf/ijcai/JoachimsSS18}, is based on the common observation \cite{DBLP:journals/tois/JoachimsGPHRG07} that users tend to pay more attention and are more likely to interact with items at the top of a recommendation list, regardless of their actual relevance or quality. Therefore, such biased interaction data may mislead newly trained models, resulting in unfairness to other items.

Traditional methods mitigate these pre-existing biases by resampling or imposing additional constraints during training \cite{DBLP:conf/www/WuCSHWW21}. However, without understanding the underlying causal mechanisms through how biased data affect the recommendation process, these methods may achieve only superficial fairness and exhibit limited generalization ability. To address this, causality-oriented methods have been proposed, analyzing and measuring the causal relationships within a biased RS. For user-side bias, counterfactual reasoning offers an elegant solution, ensuring counterfactual fairness \cite{DBLP:conf/sigir/LiCXGZ21} for individuals with sensitive attributes while generating reasonable samples \cite{DBLP:conf/sigir/MuLZWDW22} for minority users. For item-side bias, existing work often formulates SCM for thorough analysis and mitigates the negative effects through causal interventions \cite{DBLP:conf/sigir/ZhangF0WSL021,DBLP:conf/kdd/WeiFCWYH21}. The details of causal learning methods are introduced in Section \ref{sec::causal_methods}.

\subsubsection{Robustness Challenges in Data Preparation}
Since raw data are usually collected from uncontrollable environments, it is pivotal to handle the robustness challenges stemming from non-randomness \cite{DBLP:conf/www/LiXZ023} and corruption \cite{DBLP:conf/wsdm/ZhangYCHNC22} during data preparation. 

Due to the inherent high sparsity of user-item interactions, most existing work assumes that interactions are Missing At Random (MAR) for modeling simplicity and statistical convenience. Let $\mathbf{O}\in\{0,1\}^{M\times N}$ be an indicator matrix where each entry $o_{ui}=1$ if $r_{ui}$ is observed, and $o_{ui}=0$ otherwise. The MAR assumption implies that the probability of observing an interaction is independent of the unobserved rating $r_{ui}$. However, this assumption does not always hold in practice. In cases where samples are Missing Not At Random (MNAR), i.e., the missingness depends on the unobserved ratings:
\begin{equation}
P(o_{ui} = 1 \mid r_{ui}, \mathbf{X}_{ui}) \neq P(o_{ui} = 1 \mid \mathbf{X}_{ui}),
\end{equation}
where $\mathbf{X}_{ui}$ represents observed features related to user $u$ and item $i$, \textbf{selection bias} \cite{DBLP:conf/icml/LiXZ0023} becomes a significant issue. It refers to the distortion that data used to train RS models are not representative of the entire range of user preferences. To overcome this challenge, earlier work \cite{DBLP:conf/icml/Hernandez-LobatoHG14a} managed to model the generative and missing process of data, but this led to \textit{high computational costs}. Recently, propensity reweighting tools have been widely adopted to solve selection bias in RS. These methods offer higher efficiency and flexibility to integrate with other recommendation frameworks \cite{DBLP:conf/www/LiXZ023}.

Another critical robustness challenge at this stage is defending against \textbf{poisoning attacks} \cite{DBLP:conf/wsdm/ZhangYCHNC22}. These attacks usually insert fake users into training data to manipulate RS outcomes for malicious purposes. To counteract such threats, two representative defense strategies are primarily adopted, i.e., detection-based methods \cite{DBLP:conf/sigir/ZhangYCHHC20} and adversarial training \cite{DBLP:journals/tkde/TangDHYTC20}. Detection-based methods focus on developing an accurate classifier to identify and filter out spurious samples. Adversarial training, on the other hand, aims to enhance model robustness by introducing generated perturbations during training. However, the generalization ability of these defense methods is constrained by the given training data, whereas examining how attacks manipulate RS results from a causal perspective can help improve it. Useful causal techniques include counterfactual augmentation for adversarial training and causal effect inference for causality measurement.

\subsection{Challenges in Representation Learning}
Representation learning aims to learn informative user and item representations $\mathbf{h}_u$ and $\mathbf{h}_i$ from interaction data, bridging the stages of data preparation and recommendation generation. Although numerous advanced representation learning methods have been developed in RS, such as recurrent neural networks, graph neural networks \cite{DBLP:conf/aaai/MaMZSLC20} and transformer-based models \cite{DBLP:conf/cikm/SunLWPLOJ19}, they commonly overlook potential risks regarding fairness and robustness.

\subsubsection{Fairness Challenges in Representation Learning}
A major fairness challenge is how to address \textbf{conformity bias} \cite{DBLP:conf/www/ZhengGLHLJ21}. This bias is closely related to popularity bias, but specifically focuses on the perspective of user intentions. In general, a user-item interaction is usually caused by two factors: users' genuine interests $I_{\rm int}$ in the item and their conformity $I_{\rm con}$ to the item's popularity, i.e., $I_{\rm int} \rightarrow r_{ui} \leftarrow I_{\rm con}$. An unbiased model should focus on learning users' true interests rather than being misled by their conformity. However, it is intractable to clearly separate conformity intention from user representation, especially relying solely on \textit{statistical approaches}. To tackle this challenge, Zheng \textit{et al.} \shortcite{DBLP:conf/www/ZhengGLHLJ21} introduced a causality-driven framework, which leverages the colliding effect \cite{Pearl2009Causality} to learn intention-disentangled item and user representations $\{\mathbf{h}_u^{\rm int},\mathbf{h}_u^{\rm con}\}$ and $\{\mathbf{h}_i^{\rm int},\mathbf{h}_i^{\rm con}\}$ from cause-specific data.

\subsubsection{Robustness Challenges in Representation Learning}
Underlying noise is a key robust challenge in representation learning. Here, we focus on two types: spurious correlations \cite{DBLP:journals/tois/00010FSYL023} and noisy feedback \cite{DBLP:conf/sigir/ZhangJSWXW21}. 

\textbf{Spurious correlation} refers to a correlation that is statistically strong but in fact is causally incorrect. To clarify this concept, consider a straightforward example. It is a well-established fact that an item's quality influences both its price and the users' preference for it. Commonly, one can observe that higher-quality items are more expensive and more appealing to users. However, if the factor of quality is neglected, an RS may falsely imply that user preference increases along with the price. Such spurious correlations can be largely misleading. Unfortunately, \textit{statistical methods}, which typically rely on correlations or associations \cite{DBLP:journals/corr/abs-2303-11666} among samples, can hardly identify and eliminate such noise. In contrast, causality-oriented methods excel in tackling this issue by modeling and adjusting underlying causal relationships. For instance, recognizing that spurious correlations often stem from confounders, a study by \cite{DBLP:journals/corr/abs-2205-07499} performs structure adjustment to mitigate corresponding confounder effects. Another representative research by \cite{DBLP:conf/sigir/MuLZWDW22} introduces a counterfactual-based method to reduce spurious correlations by generating higher-quality counterfactual interactions. 

\textbf{Noisy feedback} manifests in two forms: 1) positive interactions from users who ultimately turned out to be unsatisfied \cite{DBLP:conf/sigir/WangF0ZC21,DBLP:conf/wsdm/WangF0NC21}; and 2) delayed interactions from users who are actually satisfied \cite{DBLP:conf/sigir/ZhangJSWXW21}. Such noisy feedback limits the ability of an RS to capture users' genuine preferences. To defend against this noise, Zhang \textit{et al.} \shortcite{DBLP:conf/sigir/ZhangJSWXW21} devised a counterfactual modifier in a reinforcement learning framework for streaming recommendation. It adopts importance sampling \cite{DBLP:conf/www/YasuiMFS20} and modifies the weight of reward in a counterfactual world to alleviate the impact of delayed feedback.

\subsection{Challenges in Recommendation Generation}
Upon obtaining representations from input data, RS models are devised to generate accurate suggestions $\hat{r}_{ui} = f(\mathbf{h}_u, \mathbf{h}_i \mid \Theta)$ that are tailored to individual preferences. Despite the promising performance that modern RS models \cite{DBLP:conf/www/LinTHZ22} achieved, there still exist several challenges regarding fairness and explainability due to their intrinsic mechanisms.

\subsubsection{Fairness Challenges in Recommendation Generation}
As RS models are commonly trained dynamically through successive feedback loops \cite{DBLP:conf/cikm/MansouryAPMB20}, biases existing in results and user feedback will be continuously amplified, known as \textbf{bias amplification}. By analyzing underlying causal relationships, Wang \textit{et al.} \shortcite{DBLP:conf/kdd/WangF0WC21} attribute bias amplification to the confounding effect arising from users' imbalanced historical distribution. Hence, they perform backdoor adjustment \cite{Pearl2009Causality} to remove causal paths from confounders, thus mitigating the impact of this issue.

Additionally, feedback loops in RS often reinforce users' existing preferences, creating an echo chamber that isolates them from diverse contents, known as the \textbf{filter bubble bias} \cite{DBLP:journals/tois/GaoWLCHLLZJ24}. To counter this challenge, a framework based on counterfactual inference \cite{DBLP:conf/sigir/WangFNC22} has been developed. This approach allows users to determine whether actions are needed to modify user embeddings through counterfactual modification. By doing so, it facilitates access to a wider and more diverse range of content, effectively helping users to break out of filter bubbles.

\subsubsection{Explainability Challenges in Recommendation}
The advancement of machine learning-based, especially deep learning-based techniques, has greatly enhanced the capability of RS models over the past decades. However, the inherent drawback of lacking interpretability has garnered significant concerns in recent years. With the introduction of complex black-box modules, it is challenging to explain how and why the recommendations are generated. This limitation not only impedes researchers in comprehensively understanding and improving models, but also undermines stakeholders' trust in RS, especially in some safety-critical applications. Therefore, effective explainers are required to provide insightful explanations for both recommendation mechanisms and results.

Despite several attempts \cite{DBLP:journals/ftir/ZhangC20} to generate explanations based on tags \cite{DBLP:conf/iui/VigSR09} or visual attentions \cite{DBLP:conf/sigir/ChenCXZ0QZ19}, they are not convincing enough and often struggle to explain the \textit{mechanism of RS models}. To bridge this gap, causality has been explored in RS to generate insightful explanations. For instance, a reinforcement learning model \cite{DBLP:conf/sigir/XianFMMZ19} is designed to discover reasonable causal paths as explanations for how recommendations are generated. However, such path-based explanations also raise concerns regarding privacy leakage. Another representative work \cite{DBLP:conf/cikm/TanXG00Z21} introduces a model-agnostic explainer based on counterfactual inference. It aims to find straightforward explanations by exploring how minimal changes $\Delta=\{\delta_1,\delta_2,\cdots,\delta_R\}$ in $R$ item features can alter recommendation outcomes $\hat{r}^\Delta_{ui}$. This counterfactual explanation not only interprets recommendation mechanisms but also has been proven to enhance recommendation performance by filtering out negative items \cite{DBLP:journals/tkde/WangLYLX24}.

\subsection{Challenges in Evaluation}
\label{sec::evaluation}
Current accuracy-oriented evaluation metrics usually concentrate on the interaction rate of the recommendation results. However, considering the fact that users are free to interact with items they want regardless of the results, an RS should focus more on the uplift of interactions that are actually brought by recommendations. This poses a significant challenge, as it is intractable to observe the interaction outcomes with and without recommendations at the same time. To address this issue, Sato \textit{et al.} \shortcite{DBLP:conf/recsys/SatoSTSZO19} proposed a causal inference framework for RS. It leverages additional interaction data of an existing model to estimate the uplift effects of a new model. Moreover, it presents a novel optimization algorithm to promote the uplift of interactions during training.

Moving beyond accuracy, TRS encompasses a variety of trustworthiness objectives, posing the necessity of specialized evaluation metrics. In particular, fairness metrics \cite{DBLP:journals/tist/LiCXGTLZ23} typically measure group or individual disparities. For instance, Wang \textit{et al.} \shortcite{DBLP:conf/sigir/WangFNC22} employed an isolation index to measure the segregation across different user groups and to assess the severity of filter bubbles. In evaluating robustness, most works \cite{DBLP:conf/www/LiXZ023,DBLP:conf/icml/LiXZ0023} test RS accuracy under varying levels of robustness challenges. In evaluating explainability, user reviews \cite{DBLP:conf/cikm/TanXG00Z21} are usually collected to assess the consistency with generated explanations. Moreover, based on the concepts of necessity and sufficiency, Tan \textit{et al.} \shortcite{DBLP:conf/cikm/TanXG00Z21} introduced two novel metrics to provide a thorough evaluation of explanation quality.

However, it should be pointed out that only a few attempts \cite{DBLP:conf/recsys/SatoSTSZO19} have been made to devise evaluations from a causal learning perspective, which is especially valuable when evaluation is required in unobservable cases, such as counterfactual scenarios. This remains an important yet underexplored direction and deserves greater research attention. Moreover, we keep in mind that the trustworthiness of RS is a human-centered concern. Therefore, beyond the technical aspects, evaluating user perception and ethical considerations \cite{TRS_TIST} is equally vital. This may require user studies to assess the subjective perceptions of fairness, robustness, explainability, and even potential societal impacts.

\begin{figure}[t]
  \centerline{\includegraphics[width=0.43\textwidth]{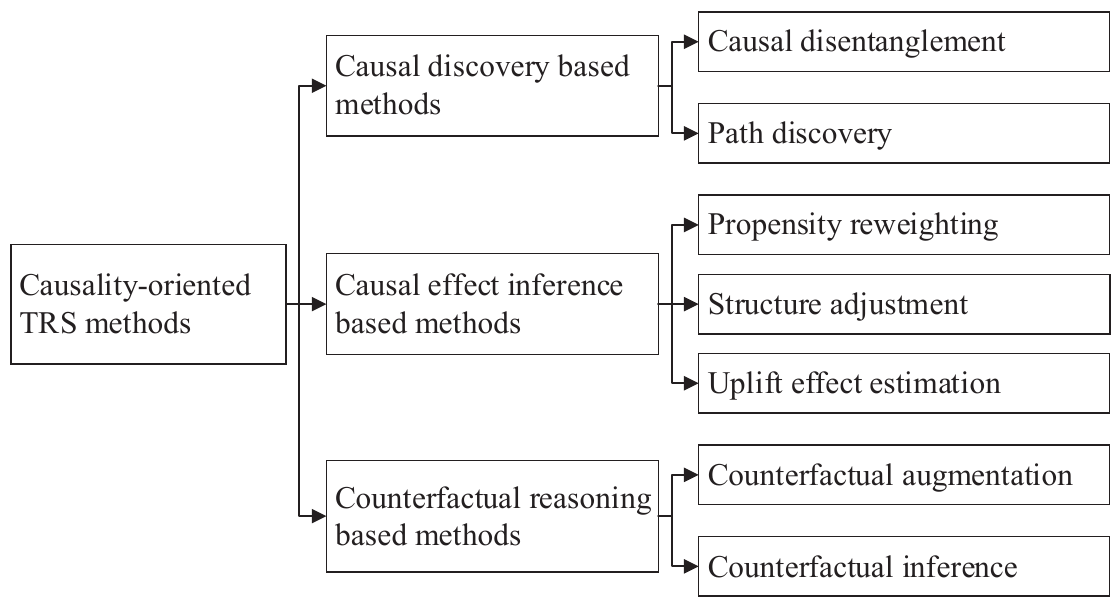}}\vspace{-4pt}
  \caption{Classification of CTRS methods.}
  \label{fig::CTRS_classification}\vspace{-10pt}
\end{figure}

\begin{table*}[ht]
\centering
\scriptsize
\renewcommand{\arraystretch}{0.95}
\begin{tabularx}{\textwidth}{>{\hsize=0.8\hsize}X|>{\hsize=0.8\hsize}X|>{\hsize=1.1\hsize}X|>{\hsize=1.3\hsize}X|>{\hsize=1.3\hsize}X|>{\hsize=0.7\hsize}X}
  \toprule[1pt]
  \multicolumn{2}{c|}{Approach} &\multicolumn{1}{c|}{Applicable challenge} &\multicolumn{1}{c|}{Pros} &\multicolumn{1}{c|}{Cons} &\multicolumn{1}{c}{Example} \\ \midrule
  \multirow{2}{=}[-16pt]{Causal discovery} &\multirow{1}{=}[-5pt]{Causal disentanglement} &\multirow{1}{=}[-8pt]{Conformity bias} &An effective way to separate interaction data based on their primary causes &Relies on prior knowledge of causal models and is hard to scale to complex scenarios &\multirow{1}{=}[-8pt]{\scriptsize [1]} \\ \cmidrule{2-6}
  &\multirow{1}{=}[-4pt]{Path discovery} &\multirow{1}{=}[-4pt]{Black-box mechanism} &Straightforward to illustrate the process of recommendations &\multirow{1}{=}[0pt]{Potential risks of privacy leakage} &\multirow{1}{=}[-4pt]{\scriptsize [2]} \\ \midrule
  \multirow{3}{=}[-25pt]{Causal effect inference} &\multirow{1}{=}[0pt]{Propensity reweighting} &\multirow{1}{=}[-4pt]{Position bias, selection bias} &\multirow{1}{=}[0pt]{Proven to be statistically unbiased} &Hard to accurately and robustly estimate the propensity weights &\multirow{1}{=}[-4pt]{\scriptsize [3], [4], [5], [6]} \\ \cmidrule{2-6}
  &\multirow{1}{=}[-4pt]{Structure adjustment} &Popularity bias, spurious correlations, bias amplification &Effective in deconfounding and estimating accurate causal effects &\multirow{1}{=}[-8pt]{Relies on given causal models} &\multirow{1}{=}[-8pt]{\scriptsize [7], [8], [9], [10]} \\ \cmidrule{2-6}
  &\multirow{1}{=}[-4pt]{Uplift effect estimation} &\multirow{1}{=}[-8pt]{Uplift evaluation} &\multirow{1}{=}[0pt]{Insightful to show the actual promotion brought by recommendations} &Additional training data are required, which can be difficult to obtain in real-world applications &\multirow{1}{=}[-8pt]{\scriptsize [11]} \\ \midrule
  \multirow{2}{=}[-22pt]{Counterfactual reasoning} &\multirow{1}{=}[-4pt]{Counterfactual augmentation} &Population bias, poisoning attacks, spurious correlations &\multirow{1}{=}[-4pt]{Excels in generating sufficient samples for training} &\multirow{1}{=}[-4pt]{Lack of assurance of the quality of generated samples} &\multirow{1}{=}[-8pt]{\scriptsize [12], [13]} \\ \cmidrule{2-6}
  &\multirow{1}{=}[-8pt]{Counterfactual inference} &Attribute bias, popularity bias, noisy feedback, filter bubble bias, unexplainable results &\multirow{1}{=}[-4pt]{A feasible way to explore causal effects of hypothetical interventions} &\multirow{1}{=}[-4pt]{Assumptions involved in computation may not always hold in practice} &\multirow{1}{=}[-4pt]{\scriptsize [14], [15], [16], [17], [18], [19], [20], [21], [22]} \\ \midrule
  \multicolumn{6}{>{\hsize=\dimexpr 6\hsize+12\tabcolsep+\arrayrulewidth\relax}X}{\scriptsize \textsuperscript{[1]}\cite{DBLP:conf/www/ZhengGLHLJ21}; \textsuperscript{[2]}\cite{DBLP:conf/sigir/XianFMMZ19}; \textsuperscript{[3]}\cite{DBLP:conf/ijcai/JoachimsSS18}; \textsuperscript{[4]}\cite{DBLP:conf/nips/Liu0023}; \textsuperscript{[5]}\cite{DBLP:conf/icml/LiXZ0023}; \textsuperscript{[6]}\cite{DBLP:conf/www/LiXZ023}; \textsuperscript{[7]}\cite{DBLP:conf/sigir/ZhangF0WSL021}; \textsuperscript{[8]}\cite{DBLP:conf/cikm/GuptaSMVS21}; \textsuperscript{[9]}\cite{DBLP:journals/tois/00010FSYL023}; \textsuperscript{[10]}\cite{DBLP:conf/kdd/WangF0WC21}; \textsuperscript{[11]}\cite{DBLP:conf/recsys/SatoSTSZO19}; \textsuperscript{[12]}\cite{DBLP:conf/sigir/MuLZWDW22}; \textsuperscript{[13]}\cite{DBLP:conf/ictir/XuGLFCZ23}; \textsuperscript{[14]}\cite{DBLP:conf/sigir/LiCXGZ21}; \textsuperscript{[15]}\cite{DBLP:conf/kdd/WeiFCWYH21}; \textsuperscript{[16]}\cite{DBLP:conf/sigir/ZhangJSWXW21}; \textsuperscript{[17]}\cite{DBLP:conf/sigir/WangFNC22}; \textsuperscript{[18]}\cite{DBLP:journals/tois/GaoWLCHLLZJ24}; \textsuperscript{[19]}\cite{DBLP:conf/cikm/TanXG00Z21}; \textsuperscript{[20]}\cite{DBLP:journals/tois/WangLYLX24}; \textsuperscript{[21]}\cite{DBLP:journals/tkde/WangLYLX24}; 
  \textsuperscript{[22]}\cite{DBLP:journals/tois/ChenWZZHXX24}}\\
  \bottomrule[1pt]
\end{tabularx}
\caption{A comparison of different classes of CTRS methods.}
\label{tab::comparison}\vspace{-10pt}
\end{table*}

\section{Causality-oriented TRS Methods}
\label{sec::causal_methods}
Upon reviewing the trustworthiness challenges at each stage of TRS and the necessity of causal solutions, we now examine various CTRS methods. First, we present a taxonomy of existing CTRS methods, as illustrated in Figure \ref{fig::CTRS_classification}, categorized based on the causal learning techniques they employ. Then, we discuss their technical details and compare different classes of CTRS methods in Table \ref{tab::comparison}, regarding their applicable challenges, pros and cons, and representative works.

\subsection{Causal Discovery based Methods}
Causal discovery, as a fundamental aspect of causal learning, identifies causal relationships from data. Unlike traditional methods that emphasize correlation \cite{DBLP:journals/corr/abs-2303-11666,DBLP:journals/corr/abs-2208-12397}, it uncovers the causality among variables and formulates causal models. This is particularly valuable in RS, where understanding the causal relationships between variables (e.g., users' intention or items' popularity) and outcomes (e.g., predicted interactions $\hat{r}_{ui}$) is essential yet intractable \cite{DBLP:conf/sigir/WangCJY23}. Two key methods have been explored for RS data: Causal Disentanglement based RS (CDRS) and Path Discovery based RS (PDRS).

\subsubsection{Causal Disentanglement based RS (CDRS)}
Underlying factors (e.g., user intentions) drive various trustworthiness issues (e.g., popularity bias), yet they are difficult to disentangle, hindering the learning of representations unaffected by them. CDRS methods \cite{DBLP:conf/www/ZhengGLHLJ21} tackle this by leveraging causality-based rules to distill cause-specific data from raw training data. For instance, based on the observation that both interest and conformity intentions affect interactions in a collider structure, $I_{\rm int} \rightarrow r_{ui} \leftarrow I_{\rm con}$, we can identify cause-specific data via the colliding effect \cite{Pearl2009Causality}. This effect implies that knowing $r_{ui}$ makes $I_{\rm int}$ and $I_{\rm con}$ dependent; for example, interactions with lower $I_{\rm int}$ are more likely to be caused by higher $I_{\rm con}$. In this way, causally disentangled representations can be learned from the refined data, effectively reducing the impact of conformity bias. However, the flexibility and applicability of CDRS can be limited due to their heavy reliance on prior knowledge of causal models and the complexity of practical scenarios.

\subsubsection{Path Discovery based RS (PDRS)}
Recent years have witnessed an upsurge in leveraging knowledge graphs \cite{DBLP:conf/sigir/YangHXL22} for accurate recommendations. To explore the causality during knowledge graph reasoning and to enhance the understanding of the recommendation mechanism, PDRS \cite{DBLP:conf/sigir/XianFMMZ19} trains a reinforcement learning agent to learn the navigation to suitable items for each user. The explored path can be used as a straightforward explanation of how recommendations are generated.

\subsection{Causal Effect Inference based Methods}
Based on the discovered or prior causality in an RS, causal effect inference aims to understand and quantify the effect of one variable on another. Generally, causal effect inference based methods can be grouped into Propensity Reweighting based RS (PRRS), Structure Adjustment based RS (SARS), and Uplift Effect Estimation based RS (UEERS).

\subsubsection{Propensity Reweighting based RS (PRRS)}
Propensity reweighting is designed to adjust the biased observations when the random experimental assignment is impractical (e.g., the cases of position bias \cite{DBLP:conf/ijcai/JoachimsSS18} and selection bias \cite{DBLP:conf/icml/LiXZ0023}). PRRS methods usually involve three key steps, i.e., propensity weight estimation, weight assignment and unbiased recommendation. First, they estimate the propensity score $\hat{p}_{ui}$, which represents the probability of receiving treatment; for example, $\hat{p}_{ui}=P(o_{ui}=1)$ represents the likelihood of observing the true rating $r_{ui}$ in MNAR cases. This estimation is usually performed by statistical methods like Naive Bayes or neural networks such as multi-layer perceptron \cite{DBLP:conf/nips/Liu0023}. Then, $\hat{p}_{ui}$ is used to inversely weight the original prediction objective $\mathcal{L}$, e.g.,
\begin{equation}
  \scalebox{0.97}{$
  \mathcal{L}' = \sum_{(u,i)\in\mathcal{D}}\frac{1}{\hat{p}_{ui}}\mathcal{L}(f(\mathbf{h}_u, \mathbf{h}_i \mid \Theta), r_{ui})
  $}.
\end{equation}
Finally, the reweighted training process helps mitigate biases, leading to more robust and fair recommendations. However, PRRS methods commonly face challenges with accuracy in weight estimation and high variance due to inverse weighting. Fortunately, recent advancements, such as IPS-V2 \cite{DBLP:conf/icml/LiXZ0023} and doubly robust model \cite{DBLP:conf/www/ZhangBLYLWR20}, have been developed to tackle these issues.

\subsubsection{Structure Adjustment based RS (SARS)}
In a given system where the underlying causal relationships are modeled by an SCM, structure adjustment emerges as a pivotal technique to perform causal interventions, e.g., mitigating the influence of confounders. Recognizing that confounders are often the primary sources of biases and noise, various structure adjustment methods have gained great attention in building TRS, including backdoor adjustment \cite{DBLP:conf/sigir/ZhangF0WSL021} and frontdoor adjustment \cite{DBLP:journals/corr/abs-2110-07122}. Backdoor adjustment utilizes the backdoor criterion \cite{Pearl2009Causality} to identify a set of observed variables $\mathbf{Z}$ that block all backdoor paths between the treatment $T$ and outcome $Y$. The causal effect is estimated by adjusting for $\mathbf{Z}$:
\begin{equation}
  \scalebox{0.94}{$
  P(Y \mid \text{do}(T)) = \sum_{\mathbf{Z}} P(Y \mid T, \mathbf{Z}) P(\mathbf{Z})
  $}.
\end{equation}
In RS, this method is applied to control for confounders like item popularity \cite{DBLP:conf/sigir/ZhangF0WSL021}, which affects both item and interaction. Frontdoor adjustment applies when confounders are unobservable, but there exists a mediator $\mathbf{M}$ that transmits the causal effect from $T$ to $Y$. Using the frontdoor criterion \cite{Pearl2009Causality}, the causal effect is computed as:
\begin{equation}
  \scalebox{0.9}{$
  P(Y \mid \text{do}(T)) = \sum_{\mathbf{M}} P(\mathbf{M} \mid T) \sum_{T'} P(Y \mid T', \mathbf{M}) P(T')
  $}.
\end{equation}

\subsubsection{Uplift Effect Estimation based RS (UEERS)}
Uplift in RS refers to the increase in user actions that are genuinely caused by recommendation outcomes. Normally, assessing these uplift effects requires knowing whether the user would interact with the item if it were recommended or not, which is, however, unfeasible to observe in practice. To handle this, Sato \textit{et al.} \shortcite{DBLP:conf/recsys/SatoSTSZO19} present an uplift evaluation based on additional interaction data observed from an existing RS model. It estimates the uplift effect of a newly deployed model by calculating the average treatment effect and leverages propensity reweighting tools to mitigate the bias arising from the non-randomness of observed data. Despite its effectiveness, the requirement for additional data in UEERS can be costly and challenging, leading to limited practicality.

\subsection{Counterfactual Reasoning based Methods}
Moving beyond causal effect inference, counterfactual reasoning focuses on discovering the effects of interventions in hypothetical scenarios to address the ``\textit{what if}'' questions. Derived from this concept, two influential tools have been widely used in CTRS: Counterfactual Augmentation based RS (CARS) and Counterfactual Inference based RS (CIRS).

\subsubsection{Counterfactual Augmentation based RS (CARS)}
Counterfactual augmentation refers to the process of enriching data within counterfactual scenarios. This usually involves intervened synthetic data points that represent plausible alternative outcomes under different conditions. For example, Zhang \textit{et al.} \shortcite{DBLP:conf/sigir/ZhangYZC021} introduced a module for this purpose to address missing data issues. It generates positive and negative samples by replacing non-essential features and essential features, respectively. By incorporating counterfactual samples, these CARS methods can be better equipped to handle diverse and unseen biases, noise, and attacks, leading to reliable recommendations in practice. However, there remains a need for effective strategies to assess and ensure the quality of counterfactual samples.

\subsubsection{Counterfactual Inference based RS (CIRS)}
Counterfactual inference is commonly used to evaluate causal relationships, aiming to answer specific ``\textit{what if}'' questions. These questions vary across different problems. For instance, to handle the popularity bias in RS, Wei \textit{et al.} \shortcite{DBLP:conf/kdd/WeiFCWYH21} investigated ``\textit{what would the ranking score be if the model only uses item properties}''. By analyzing the recommendation mechanism with counterfactual cases, they identified that the key to debiasing is reducing the direct effect of popularity $I_{\rm pop}$ on the outcome $Y$ and focusing on the total indirect effect \cite{Pearl2009Causality}. However, CIRS methods often involve assumptions to estimate counterfactual results, which may not always hold in practice, potentially limiting their generalization ability.

\section{Open Problems and Future Directions}

\textbf{Unresolved challenges in fairness and robustness.}
As shown in Table \ref{tab::classification}, several critical trustworthiness challenges, such as population bias and poisoning attacks, remain insufficiently addressed. These issues lead to unfair performance disparities and increased vulnerability to malicious attacks. Thus, there is a clear need for targeted research from a causal perspective to analyze and develop tailored solutions.

\textbf{Beyond causal prior knowledge.}
Most existing CTRS \cite{DBLP:journals/tois/00010FSYL023} rely on handcrafted causal analysis and predefined causal models, limiting their effectiveness in complex cases. While causal discovery has developed advanced techniques (e.g., $\rm A^*$) \cite{DBLP:conf/aaai/LuZY21} in causal learning, it remains underexplored in CTRS. Incorporating these methods is crucial for improved practicality and applicability.

\textbf{Causality-oriented trustworthy evaluations.}
As discussed in Section \ref{sec::evaluation}, research on evaluating RS from causal learning perspectives \cite{DBLP:conf/recsys/SatoSTSZO19} is limited. Advancing causal-oriented evaluations can bridge the gap between causal theory and practical RS assessment. Moreover, ensuring trustworthiness across all stages is crucial, highlighting the necessity of multi-stage trustworthy evaluations.


\textbf{Causality-oriented trustworthy generative RS.}
The integration of generative models (e.g., diffusion) into RS \cite{Li2025Generating,DBLP:conf/nips/YangWWWY023} introduces new trustworthiness challenges, including the risk of generating biased and uninterpretable recommendation content. Future research should develop causal generative frameworks that apply causal inference to intervene in the biased generation process and leverage counterfactual reasoning for improved interpretability.

\textbf{Causality enhanced LLM-based RS.}
Large Language Models (LLMs) have been successfully introduced to RS \cite{DBLP:conf/sigir/LinWLYFWC24}, enhancing reasoning capabilities but also raising concerns about inherited stereotypes and bias \cite{DBLP:conf/ci2/KotekDS23} in LLMs. Research directions include using causal inference in fine-tuning to reduce bias propagation.

\section{Conclusions}
Recommender systems are integral to daily life. Growing concerns over trustworthiness have led to Causality-oriented Trustworthy RS (CTRS), which leverages causal learning to enhance fairness, robustness, and explainability. However, there are no dedicated surveys devoted to this area. 
This survey presents a comprehensive overview of CTRS, identifying key trustworthiness challenges across the RS lifecycle and connecting them to the corresponding causal learning based solutions. A comprehensive summarization and comparison of CTRS methods have been provided. In addition, we compile representative open-source CTRS algorithms and datasets at \url{https://github.com/JinLi-i/CTRS_resource}.




\clearpage
\bibliographystyle{named}
\bibliography{mybibtex}

\end{document}